# MEASUREMENTS OF THE RATIOS $\mathcal{B}(D_s^+ \to \eta \ell^+ \nu)/\mathcal{B}(D_s^+ \to \phi \ell^+ \nu)$ AND $\mathcal{B}(D_s^+ \to \eta' \ell^+ \nu)/\mathcal{B}(D_s^+ \to \phi \ell^+ \nu)$.

G. Brandenburg,[1] D. Cinabro,[1] T. Liu,[1] M. Saulnier,[1] R. Wilson,[1] H. Yamamoto,[1] T. Bergfeld,[2] B.I. Eisenstein,[2] J. Ernst,[2] G.E. Gladding,[2] G.D. Gollin,[2] M. Palmer,[2] M. Selen,[2] J.J. Thaler,[2] K.W. Edwards,[3] K.W. McLean,[3] M. Ogg,[3] A. Bellerive,[4] D.I. Britton,[4] E.R.F. Hyatt,[4] R. Janicek,[4] D.B. MacFarlane,[4] P.M. Patel,[4] B. Spaan,[4] A.J. Sadoff,[5] R. Ammar,[6] P. Baringer,[6] A. Bean,[6] D. Besson,[6] D. Coppage,[6] N. Copty,[6] R. Davis,[6] N. Hancock,[6] S. Kotov,[6] I. Kravchenko,[6] N. Kwak,[6] Y. Kubota,[7] M. Lattery,[7] M. Momayezi,[7] J.K. Nelson,[7] S. Patton,[7] R. Poling,[7] V. Savinov,[7] S. Schrenk,[7] R. Wang,[7] M.S. Alam,[8] I.J. Kim,[8] Z. Ling,[8] A.H. Mahmood,[8] J.J. O'Neill,[8] H. Severini,[8] C.R. Sun,[8] F. Wappler,[8] G. Crawford,[9] J.E. Duboscq,[9] R. Fulton,[9] D. Fujino,[9] K.K. Gan,[9] K. Honscheid,[9] H. Kagan,[9] R. Kass,[9] J. Lee,[9] M. Sung,[9] C. White,[9] A. Wolf,[9] M.M. Zoeller,[9] X. Fu,[10] B. Nemati,[10] W.R. Ross,[10] P. Skubic,[10] M. Wood,[10] M. Bishai,[11] J. Fast,[11] E. Gerndt,[11] J.W. Hinson,[11] T. Miao,[11] D.H. Miller,[11] M. Modesitt,[11] E.I. Shibata,[11] I.P.J. Shipsey,[11] P.N. Wang,[11] L. Gibbons,[12] S.D. Johnson,[12] Y. Kwon,[12] S. Roberts,[12] E.H. Thorndike,[12] T.E. Coan,[13] J. Dominick,[13] V. Fadeyev,[13] I. Korolkov,[13] M. Lambrecht,[13] S. Sanghera,[13] V. Shelkov,[13] T. Skwarnicki,[13] R. Stroynowski,[13] I. Volobouev,[13] G. Wei,[13] M. Artuso,[14] M. Gao,[14] M. Goldberg,[14] D. He,[14] N. Horwitz,[14] S. Kopp,[14] G.C. Moneti,[14] R. Mountain,[14] F. Muheim,[14] Y. Mukhin,[14] S. Playfer,[14] S. Stone,[14] X. Xing,[14] J. Bartelt,[15] S.E. Csorna,[15] V. Jain,[15] S. Marka,[15] D. Gibaut,[16] K. Kinoshita,[16] P. Pomianowski,[16] B. Barish,[17] M. Chadha,[17] S. Chan,[17] D.F. Cowen,[17] G. Eigen,[17] J.S. Miller,[17] C. O'Grady,[17] J. Urheim,[17] A.J. Weinstein,[17] F. Würthwein,[17] D.M. Asner,[18] M. Athanas,[18] D.W. Bliss,[18] W.S. Brower,[18] G. Masek,[18] H.P. Paar,[18] J. Gronberg,[19] C.M. Korte,[19] R. Kutschke,[19] S. Menary,[19] R.J. Morrison,[19] S. Nakanishi,[19] H.N. Nelson,[19] T.K. Nelson,[19] C. Qiao,[19] J.D. Richman,[19] D. Roberts,[19] A. Ryd,[19] H. Tajima,[19] M.S. Witherell,[19] R. Balest,[20] K. Cho,[20] W.T. Ford,[20] M. Lohner,[20] H. Park,[20] P. Rankin,[20] J.G. Smith,[20] J.P. Alexander,[21] C. Bebek,[21] B.E. Berger,[21] K. Berkelman,[21] K. Bloom,[21] T.E. Browder,[21*] D.G. Cassel,[21] H.A. Cho,[21] D.M. Coffman,[21] D.S. Crowcroft,[21] M. Dickson,[21] P.S. Drell,[21] D.J. Dumas,[21] R. Ehrlich,[21] R. Elia,[21] P. Gaidarev,[21] M. Garcia-Sciveres,[21] B. Gittelman,[21] S.W. Gray,[21] D.L. Hartill,[21] B.K. Heltsley,[21] S. Henderson,[21] C.D. Jones,[21] S.L. Jones,[21] J. Kandaswamy,[21] N. Katayama,[21] P.C. Kim,[21] D.L. Kreinick,[21] T. Lee,[21] Y. Liu,[21] G.S. Ludwig,[21] J. Masui,[21] J. Mevissen,[21] N.B. Mistry,[21] C.R. Ng,[21] E. Nordberg,[21] J.R. Patterson,[21] D. Peterson,[21] D. Riley,[21] A. Soffer,[21] P. Avery,[22] A. Freyberger,[22] K. Lingel,[22] C. Prescott,[22] J. Rodriguez,[22] S. Yang,[22] and J. Yelton[22]

(CLEO Collaboration)

[1]*Harvard University, Cambridge, Massachusetts 02138*

[2]*University of Illinois, Champaign-Urbana, Illinois, 61801*

[3]*Carleton University, Ottawa, Ontario K1S 5B6 and the Institute of Particle Physics, Canada*

[4] *McGill University, Montréal, Québec H3A 2T8 and the Institute of Particle Physics, Canada*
[5] *Ithaca College, Ithaca, New York 14850*
[6] *University of Kansas, Lawrence, Kansas 66045*
[7] *University of Minnesota, Minneapolis, Minnesota 55455*
[8] *State University of New York at Albany, Albany, New York 12222*
[9] *Ohio State University, Columbus, Ohio, 43210*
[10] *University of Oklahoma, Norman, Oklahoma 73019*
[11] *Purdue University, West Lafayette, Indiana 47907*
[12] *University of Rochester, Rochester, New York 14627*
[13] *Southern Methodist University, Dallas, Texas 75275*
[14] *Syracuse University, Syracuse, New York 13244*
[15] *Vanderbilt University, Nashville, Tennessee 37235*
[16] *Virginia Polytechnic Institute and State University, Blacksburg, Virginia, 24061*
[17] *California Institute of Technology, Pasadena, California 91125*
[18] *University of California, San Diego, La Jolla, California 92093*
[19] *University of California, Santa Barbara, California 93106*
[20] *University of Colorado, Boulder, Colorado 80309-0390*
[21] *Cornell University, Ithaca, New York 14853*
[22] *University of Florida, Gainesville, Florida 32611*

## Abstract

Using the CLEO II detector we measure $\mathcal{B}(D_s^+ \to \eta e^+\nu)/\mathcal{B}(D_s^+ \to \phi e^+\nu) = 1.24 \pm 0.12 \pm 0.15$, $\mathcal{B}(D_s^+ \to \eta' e^+\nu)/\mathcal{B}(D_s^+ \to \phi e^+\nu) = 0.43 \pm 0.11 \pm 0.07$ and $\mathcal{B}(D_s^+ \to \eta' e^+\nu)/\mathcal{B}(D_s^+ \to \eta e^+\nu) = 0.35 \pm 0.09 \pm 0.07$. We find the vector to pseudoscalar ratio, $\mathcal{B}(D_s^+ \to \phi e^+\nu)/\mathcal{B}(D_s^+ \to (\eta+\eta')e^+\nu) = 0.60\pm0.06\pm0.06$, which is similar to the ratio found in non strange $D$ decays.

PACS numbers: 13.20.Fc, 13.65+i, 14.40.Lb

*Permanent address: University of Hawaii at Manoa

One of the outstanding problems in charm semileptonic decay is the disagreement on the vector to pseudoscalar ratio, $\mathcal{B}(D \to \overline{K}^{*} e^+\nu)/\mathcal{B}(D \to \overline{K} e^+\nu)$. The experimental world average of this ratio is $0.56 \pm 0.06$ [1], while theoretical predictions range from 0.5 to 1.2 [1,2]. It is of interest to repeat these measurements in the $D_s^+$ sector, where the initial and final hadrons differ by the substitution of a light quark by a strange quark. In the $D_s^+$ sector only the $D_s^+ \to \phi\ell^+\nu$ decay has been extensively studied. The Fermilab experiment E653 has seen evidence for the remaining major $D_s^+$ semileptonic modes, $D_s^+ \to (\eta + \eta')\mu^+\nu$ [3]. In this paper we report the first measurements of $\mathcal{B}(D_s^+ \to \eta e^+\nu)$ and $\mathcal{B}(D_s^+ \to \eta' e^+\nu)$ normalized to $\mathcal{B}(D_s^+ \to \phi e^+\nu)$, and the vector to pseudoscalar ratio. The data include events with both muons and electrons.

The data consists of an integrated luminosity of 3.11 fb$^{-1}$ of $e^+e^-$ collisions recorded with the CLEO II detector at the Cornell Electron Storage Ring (CESR). The data sample contains about 3.7 million $e^+e^- \to c\bar{c}$ events taken at center-of-mass energies on the $\Upsilon(4S)$ resonance and in the nearby continuum ($\sqrt{s} \sim 10.6$ GeV). The CLEO II detector includes a CsI electromagnetic calorimeter that provides excellent $\eta$ reconstruction [4].

Due to the undetected neutrino, we cannot fully reconstruct $D_s^+ \to X\ell^+\nu$ decays, where $X \equiv \phi$, $\eta$ or $\eta'$. However, there are few processes which produce both an $X$ and a lepton in the same jet. This correlation is used to extract a clean $D_s^+ \to X\ell^+\nu$ signal and is implemented by requiring that the $X$ and the lepton be in the same hemisphere with respect to the thrust axis of the event. Backgrounds can be reduced for $D_s^+ \to \eta\ell^+\nu$ and $D_s^+ \to \phi\ell^+\nu$, which have sufficient statistics, by also detecting the low energy photon from the $D_s^{*+} \to D_s^+\gamma$ decay. We refer to this second method as the $D_s^{*+}$ tag analysis.

Electron and muon candidates are restricted to lie in the fiducial regions $|\cos\theta| < 0.91$ and $|\cos\theta| < 0.81$, respectively, where $\theta$ is the polar angle of the track with respect to the beam axis. Electron candidates are required to have momenta above 0.7 GeV/c and 1.0 GeV/c for the $D_s^{*+}$ tag and non $D_s^{*+}$ tag analyses, respectively. Electrons are identified by comparing their ionization energy loss, time-of-flight, energy deposit and shower shapes in the electromagnetic calorimeter with that expected for true electrons. Electrons from photon conversions and Dalitz decays of $\pi^0$'s are rejected. Muon candidates in the region $|\cos\theta| < 0.61$ ($|\cos\theta| > 0.61$) are required to have momenta above 1.5 GeV/c (1.9 GeV/c). In order to be identified as a muon, a track must penetrate at least 5 interaction lengths of iron.

We identify $\phi$, $\eta$ and $\eta'$ candidates by detecting the decays to the $K^+K^-$, $\gamma\gamma$ and $\eta\pi^+\pi^-$ states, respectively. We require that the momenta of these states be greater than 1.0 GeV/c to reduce combinatoric background. Charged kaon (pion) candidates are required to have ionization energy loss and time-of-flight consistent with that expected for true kaons (pions). Momentum dependent efficiencies for identifying kaons in $\phi$ decays are obtained from the data by comparing the inclusive $\phi$ yield before and after particle identification. These efficiencies are then combined with the predicted $\phi$ momentum spectrum from $D_s^+ \to \phi\ell^+\nu$ Monte Carlo events to give the total $\phi$ identification efficiency. Photon candidates must lie in the fiducial region $|\cos\theta| < 0.81$ and have a lateral shower shape consistent with that expected for true photons. In reconstructing the $\eta$ we require $|\cos\theta_d| < 0.9$ to reduce random combinations of low momentum photons, where $\theta_d$ is the photon decay angle in the $\eta$ rest frame with respect to the $\eta$ direction in the laboratory. We veto any photon which, when combined with another photon, has an invariant mass consistent with the $\pi^0$ mass and a momentum greater than 0.8 GeV/c. In reconstructing the $\eta'$ we require that the invariant mass of the two photons from $\eta \to \gamma\gamma$ decays be consistent with the $\eta$ mass and the $\eta$ momentum be greater than 0.5 GeV/c. Pion identification efficiency in $\eta'$ decays is obtained from the inclusive $\eta'$ data in the same manner as the kaon identification efficiency for $\phi$ reconstruction.

To suppress the combinatoric background from $\Upsilon(4S)$ events, which tend to be more spherical, we require that the ratio of Fox-Wolfram moments [5], $R_2 = H_2/H_0$, be greater than 0.30. The $X\ell^+$ candidates must have an invariant mass less than 1.9 GeV/c$^2$. For $\eta\ell^+$ candidates we require that the $\eta\ell^+$ invariant mass be greater than 1.2 GeV/c$^2$ since backgrounds are large at low $\eta\ell^+$ mass. In addition, we require that the $X\ell^+$ momentum be less than 4.5 GeV/c, and greater than 2.0 and 2.5 GeV/c for the $D_s^{*+}$ tag and non tag analyses, respectively. Efficiencies are determined with a Monte Carlo simulation using the ISGW model [6]. These Monte Carlo events are passed through a full simulation of the CLEO II detector and the same event reconstruction and analysis chain as the data.

The $K^+K^-$, $\gamma\gamma$ and $\eta\pi^+\pi^-$ invariant mass spectra for all $X\ell^+$ combinations which pass the above selection criteria are fit to obtain the number of candidates which are given in Table I. Figures 1(a) and (b) show the invariant mass spectra for $\gamma\gamma$ and $\eta\pi^+\pi^-$.

There are two main sources of background in the sample: $X$'s accompanied by fake leptons [7], and random $X\ell^+$ combinations. The background due to fake leptons is estimated by using the real data to measure the momentum dependent probabilities that a hadron will be misidentified as a lepton. These probabilities are typically 0.3% for electrons and 1.2% for muons. To estimate the number of events due to fake leptons we take the events which pass the basic event selection cuts and treat all charged tracks in these events, which do not pass the lepton identification criteria, as leptons. The entries which pass all the cuts are weighted by the appropriate fake probability and summed to give the fake background. All background estimates and signal yields are given in Table I.

Random $X\ell^+$ combinations come from $e^+e^- \to c\bar{c}$ events in which an $X$ produced in the fragmentation process is combined with a lepton from a semileptonic decay of the charmed hadron in the same jet, and from $\Upsilon(4S) \to B\bar{B}$ events in which an $X$ and a lepton produced in the decay of the $B$ and $\bar{B}$ mesons are combined. The background from random $X\ell^+$ combinations is estimated using the Monte Carlo simulation. However, the $X$ production rate from both fragmentation and $B$ meson decays is not well simulated. For this reason the magnitudes of these backgrounds are determined from the data.

To determine the rate at which fragmentation $X$'s are produced in the same jet as a charmed meson we use a sample containing reconstructed $D^0 \to K^-\pi^+$ decays with $D^0$ momenta in the same range as those of our $D_s^+$ sample. Since leptons come primarily from charm semileptonic decays, the direction of a charmed hadron is close to that of the high momentum lepton. The yields of $D^0$ and $X$ mesons produced in the same hemisphere are obtained by fitting their invariant mass distributions. In data $1.2 \pm 0.5$ $\eta$ mesons are found for every 1000 reconstructed $D^0$ mesons, which is to be compared with $1.7 \pm 0.1$ in the $e^+e^- \to c\bar{c}$ Monte Carlo sample. Therefore for $D_s^+ \to \eta\ell^+\nu$ we scale the Monte Carlo estimate of the charm continuum background by $0.7 \pm 0.3$. The scale factors for $D_s^+ \to \phi\ell^+\nu$ and $D_s^+ \to \eta'\ell^+\nu$ are $1.0 \pm 0.7$ and $0.0^{+0.4}_{-0.0}$, respectively. With these factors applied, the Monte Carlo simulation predicts the continuum $c\bar{c}$ background.

The background from random $X\ell^+$ combinations in $B\bar{B}$ events is estimated in a similar manner. In this case the directions of the $X$ and the lepton are uncorrelated. Therefore we compare the number of $X$'s with momentum above 1.0 GeV/c in the continuum subtracted $\Upsilon(4S)$ data with the number observed in the $\Upsilon(4S)$ $B\bar{B}$ Monte Carlo sample. The resulting backgrounds are found by scaling the Monte Carlo estimate.

TABLE I. Summary of the non $D_s^{*+}$ and $D_s^{*+}$ tag analyses. The errors quoted in this table are statistical only. To get the corrected yield, we divide the signal yield by the detection efficiency ($\epsilon$) and the branching fraction ($\mathcal{B}$) for the subsequent decays of $\phi$, $\eta$ or $\eta'$ mesons, and a factor of two to take into account the fact that the lepton can be either electron or muon.

| | non $D_s^{*+}$ tag | | | $D_s^{*+}$ tag | |
|---|---|---|---|---|---|
| Decay mode | $D_s^+ \to \phi\ell^+\nu$ | $D_s^+ \to \eta\ell^+\nu$ | $D_s^+ \to \eta'\ell^+\nu$ | $D_s^+ \to \phi\ell^+\nu$ | $D_s^+ \to \eta\ell^+\nu$ |
| Candidates | $863.6 \pm 40.5$ | $577.6 \pm 30.3$ | $42.3 \pm 7.9$ | $326.9 \pm 27.1$ | $153.4 \pm 15.0$ |
| Fake lepton | $95.3 \pm 2.2$ | $128.8 \pm 1.6$ | $9.0 \pm 0.4$ | $27.9 \pm 1.2$ | $22.7 \pm 0.7$ |
| $c\bar{c} + B\bar{B}$ | $72.7 \pm 2.0$ | $62.7 \pm 3.2$ | $1.9 \pm 0.2$ | $23.5 \pm 1.0$ | $12.3 \pm 1.4$ |
| $D_s^+ \to \eta'\ell^+\nu$ | | $5.0 \pm 1.3$ | | | $2.3 \pm 0.6$ |
| $D_s^+ + D^+$ | $695.6 \pm 40.6$ | $381.0 \pm 30.5$ | $31.4 \pm 7.9$ | | |
| $f_c$ | | 0.855 | 0.926 | | 0.989 |
| $D_s^+$ signal yield | $695.6 \pm 40.6$ | $325.6 \pm 26.1$ | $29.1 \pm 7.3$ | | |
| $D_s^{*+}$ + Photon bkgd. | | | | $275.5 \pm 27.2$ | $114.8 \pm 15.1$ |
| $f_{sig}$ | | | | 0.759 | 0.803 |
| $D_s^{*+}$ signal yield | | | | $209.2 \pm 20.6$ | $92.2 \pm 12.1$ |
| $\epsilon \cdot \mathcal{B}(\%)$ | 2.078 | 0.807 | 0.204 | 1.408 | 0.474 |
| Corrected yield | $16734 \pm 976$ | $20184 \pm 1616$ | $7128 \pm 1786$ | $7431 \pm 733$ | $9730 \pm 1276$ |

For $D_s^+ \to \eta\ell^+\nu$ decay, we estimate the $D_s^+ \to \eta'\ell^+\nu$ feed down by multiplying the efficiency corrected yield of the $D_s^+ \to \eta'\ell^+\nu$ decay in our data sample, as discussed below, by the appropriate efficiency and the branching fraction. The Cabibbo suppressed $D^+ \to \eta\ell^+\nu$ decay contributes a sizable contamination to the $D_s^+ \to \eta\ell^+\nu$ decay in the non $D_s^{*+}$ tag analysis. Using a measurement [13] of the ratio of $\sigma(D^+)/\sigma(D_s^+)$ and the ISGW2 model [10] for the ratio of widths, $\Gamma(D^+ \to \eta(\eta')\ell^+\nu)/\Gamma(D_s^+ \to \eta(\eta')\ell^+\nu)$, we estimate the fraction of the yield which is from $D_s^+$ to be $f_c^{\eta}(f_c^{\eta'}) = 0.855 \pm 0.051(0.926 \pm 0.026)$ for an $\eta - \eta'$ mixing angle of $(-15 \pm 5)°$.

Figures 1(c) and (d) show the fitted number of $\eta$'s and $\eta'$'s which fall in each bin of the $\eta\ell^+$ and $\eta'\ell^+$ mass, respectively. The combined background estimate is also shown, as well as the sum of the simulated signal and the combined background estimate. The simulated signal has been normalized to the number of candidates extracted from the fit to the $\gamma\gamma$ and $\eta\pi^+\pi^-$ invariant mass spectra. Table I gives the yields after correcting for efficiencies and the $\phi \to K^+K^-$, $\eta \to \gamma\gamma$ and $\eta' \to \eta\pi^+\pi^-$ branching fractions.

In the $D_s^{*+}$ tag analysis, we look for a low energy photon from the $D_s^{*+} \to D_s^+\gamma$ decay in the same jet as the $X(\equiv \phi$ or $\eta)$ and the lepton. The photon candidate must now lie in the fiducial region $|\cos\theta| < 0.71$, and have energy greater than 0.12 GeV. To further suppress background photons from $\pi^0$ decays, we veto any photon that has an invariant mass consistent with the $\pi^0$ mass when combined with another photon. To select $X\ell^+\gamma$ candidates we require that $\Delta M \equiv M_{X\ell^+\gamma} - M_{X\ell^+}$ be between 0.1 and 0.2 GeV/c$^2$, where $M_{X\ell^+}$ and $M_{X\ell^+\gamma}$ are the invariant masses of $X\ell^+$ and $X\ell^+\gamma$ systems, respectively. Figure 2(a) and (b) show the $K^+K^-$ and $\gamma\gamma$ invariant mass spectra for all $\phi\ell^+\gamma$ and $\eta\ell^+\gamma$ combinations which pass the above selection criteria.

The backgrounds due to fake leptons, random $X\ell^+$ combinations, and the $D_s^+ \to \eta'\ell^+\nu$ feed down are estimated by the same procedures as for the non $D_s^{*+}$ tag analysis. The contamination from the Cabibbo suppressed decay $D^+ \to \eta\ell^+\nu$ is estimated to be small. The background from the combination of a true $X\ell^+$ and a random photon is estimated using a Monte Carlo simulation. To test the agreement between the Monte Carlo estimate and the data for this background, we use events with the decay chain, $D^{*+} \to D^0\pi^+$ and $D^0 \to K^-e^+\nu$. These events are ideally suited for the test because a $K^-e^+$ pair has kinematics similar to that of an $X\ell^+$ pair from the $D_s^+$ decay. We combine photons with $K^-e^+$ pairs selected with the same criteria as for $X\ell^+$'s from the $D_s^+$ decay. In data $21.4 \pm 1.7$ $K^-e^+\gamma$'s are found for every 100 reconstructed $K^-e^+$ pairs, which is to be compared with $22.3 \pm 0.1$ for the Monte Carlo sample. Therefore we scale the Monte Carlo estimate of the random photon background by the factor of $0.96 \pm 0.08$. With this scale factor, the Monte Carlo simulation predicts the ratio of the signal to the sum of the signal and the random photon background, $f_{sig}$, to be $0.759 \pm 0.030$ for $D_s^+ \to \phi\ell^+\nu$ and $0.803 \pm 0.031$ for $D_s^+ \to \eta\ell^+\nu$. Figures 2(c) and (d) show the fitted number of $\phi$'s and $\eta$'s which fall in each $\Delta M$ bin, respectively. The combined background estimate is also shown, as well as the sum of the simulated signal and the combined background estimate. The simulated signal has been normalized to the number of candidates extracted from the fit to the $K^+K^-$ and $\gamma\gamma$ invariant mass spectra.

To obtain the effective branching ratio in the electron channel, we take into account the reduced phase space for the muon channel [8], and the efficiency loss due to the final state radiation for the electron channel [9]. The results for $\mathcal{B}(D_s^+ \to \eta e^+\nu)/\mathcal{B}(D_s^+ \to \phi e^+\nu)$ are $1.21 \pm 0.12 \pm 0.16$ and $1.32 \pm 0.22 \pm 0.15$ for the nontag and tag analyses, respectively. The two measurements for $\eta/\phi$ are combined taking into account the correlated statistical and systematic errors. This result and the $\eta'/\phi$ measurement from the nontag analysis are given in Table II.

The major contributions to the systematic error in $\mathcal{B}(D_s^+ \to \eta e^+\nu)/\mathcal{B}(D_s^+ \to \phi e^+\nu)$ are the uncertainties in the $D^+ \to \eta\ell^+\nu$ contamination, fitting, fake lepton, and charm continuum backgrounds. The systematic errors in $\mathcal{B}(D_s^+ \to \eta' e^+\nu)/\mathcal{B}(D_s^+ \to \phi e^+\nu)$ and $\mathcal{B}(D_s^+ \to \eta' e^+\nu)/\mathcal{B}(D_s^+ \to \eta e^+\nu)$ are dominated by the uncertainty in the charm continuum background.

Model predictions and measurements are listed in Table II. Our measurements of $\mathcal{B}(D_s^+ \to \eta e^+\nu)/\mathcal{B}(D_s^+ \to \phi e^+\nu)$ and $\mathcal{B}(D_s^+ \to \eta' e^+\nu)/\mathcal{B}(D_s^+ \to \phi e^+\nu)$ agree well with the ISGW2 model predictions with an $\eta - \eta'$ mixing angle of $-10^\circ$ [10]. Our measurement of the vector to pseudoscalar ratio for the $D_s^+$ semileptonic decays, $0.60 \pm 0.06 \pm 0.06$, agrees well with the measurements for $D$ semileptonic decays, $0.56 \pm 0.06$ [1], and also with the ISGW2 model prediction [10]. Using the factorization hypothesis, Kamal *et al.* predict $\mathcal{B}(D_s^+ \to \eta' e^+\nu)/\mathcal{B}(D_s^+ \to \eta e^+\nu) = \mathcal{B}(D_s^+ \to \eta'\rho^+)/\mathcal{B}(D_s^+ \to \eta\rho^+)$ [11]. We obtain $\mathcal{B}(D_s^+ \to \eta'\rho^+)/\mathcal{B}(D_s^+ \to \eta\rho^+) = 1.20 \pm 0.35$ from the CLEO measurements [12]. This disagrees with our measurement of $\mathcal{B}(D_s^+ \to \eta' e^+\nu)/\mathcal{B}(D_s^+ \to \eta e^+\nu) = 0.35 \pm 0.09 \pm 0.07$ from the nontag analysis.

In conclusion, we have measured $\mathcal{B}(D_s^+ \to \eta e^+\nu)/\mathcal{B}(D_s^+ \to \phi e^+\nu) = 1.24 \pm 0.12 \pm 0.15$

TABLE II. Summary of measurements and predictions. The numbers in (parentheses) [brackets] are the ISGW2 predictions using values of the $\eta - \eta'$ mixing angle of $(-10^\circ)$ and $[-20^\circ]$.

| | This Experiment | E653 | ISGW2 |
|---|---|---|---|
| $\frac{\mathcal{B}(D_s^+ \to \eta e^+ \nu)}{\mathcal{B}(D_s^+ \to \phi e^+ \nu)}$ | $1.24 \pm 0.12 \pm 0.15$ | | (1.17)[0.77] |
| $\frac{\mathcal{B}(D_s^+ \to \eta' e^+ \nu)}{\mathcal{B}(D_s^+ \to \phi e^+ \nu)}$ | $0.43 \pm 0.11 \pm 0.07$ | $< 1.6$ @90%C.L. | (0.50)[0.67] |
| $\frac{\mathcal{B}(D_s^+ \to \phi e^+ \nu)}{\mathcal{B}(D_s^+ \to (\eta + \eta') e^+ \nu)}$ | $0.60 \pm 0.06 \pm 0.06$ | $0.26^{+0.18}_{-0.07}$ | (0.60)[0.69] |

and $\mathcal{B}(D_s^+ \to \eta' e^+ \nu)/\mathcal{B}(D_s^+ \to \phi e^+ \nu) = 0.43 \pm 0.11 \pm 0.07$. Our value for the ratio of vector to pseudoscalar rates for the $D_s^+$ semileptonic decays agrees with that observed for the $D$ semileptonic decays, and also with the ISGW2 model prediction. This agreement increases confidence in the evaluations of the $D_s^+ \to \phi \pi^+$ branching ratio using models of semileptonic decays and measurements of $\Gamma(D_s^+ \to \phi \ell^+ \nu)/\Gamma(D_s^+ \to \phi \pi^+)$ [1,2].

We gratefully acknowledge the effort of the CESR staff in providing us with excellent luminosity and running conditions. This work was supported by the National Science Foundation, the U.S. Department of Energy, the Heisenberg Foundation, the Alexander von Humboldt Stiftung, the Natural Sciences and Engineering Research Council of Canada, and the A.P. Sloan Foundation.

FIG. 1. The (a) $\gamma\gamma$ and (b) $\eta\pi^+\pi^-$ invariant mass spectra for $D_s^+ \to \eta\ell^+\nu$ and $D_s^+ \to \eta'\ell^+\nu$ candidates in the non $D_s^{*+}$ tag analysis: The solid curve is a fit to each spectrum. The (c) $\eta\ell^+$ and (d) $\eta'\ell^+$ invariant mass spectra for the candidates: The points with error bars represent the number of candidates in each mass bin. The solid histogram shows the simulated signal plus the predicted background, and the dotted histogram shows the predicted background. We removed the cut on $\eta\ell^+$ and $\eta'\ell^+$ invariant masses discussed in the text for (c) and (d).

FIG. 2. The (a) $K^+K^-$ and (b) $\gamma\gamma$ invariant mass spectra for $D_s^+ \to \phi\ell^+\nu$ and $D_s^+ \to \eta\ell^+\nu$ candidates in the $D_s^{*+}$ tag analysis: Note that a candidate photon from the $D_s^{*+}$ is required. The solid curve is a fit to each spectrum. The spectra of the pseudo mass difference $\Delta M$ for (c) $D_s^+ \to \phi\ell^+\nu$ and (d) $D_s^+ \to \eta\ell^+\nu$ candidates: The points with error bars represent the number of candidates in each $\Delta M$ bin. The solid histogram shows the simulated signal plus the predicted background, and the dotted histogram shows the predicted background. We removed the cut on $\Delta M$ discussed in the text for (c) and (d).